\documentclass[%
 reprint,
 amsmath,amssymb,
 aps,
]{revtex4-1}

\usepackage{graphicx}
\usepackage{dcolumn}
\usepackage{bm}
\usepackage{booktabs}
\usepackage{multirow}
\usepackage[normalem]{ulem}
\usepackage{subcaption}
\usepackage{pgf}
\usepackage{siunitx}

\begin{document}

\preprint{APS/123-QED}

\title{Generating Minimal Training Sets for Machine Learned Potentials}

\author{Jan Finkbeiner}\thanks{Contributed equally}
\affiliation{%
 Peter Gr\"{u}nberg Institute \\ Forschungszentrum J\"{u}lich GmbH \\ Wilhelm-Johnen-Straße, 52428 J\"{u}lich
}%
\author{Samuel Tovey$^{*}$}%
\author{Christian Holm}
 \email{holm@icp.uni-stuttgart.de}
\affiliation{%
 Institute for Computational Physics \\ University of Stuttgart \\ Allmandring 3, 70569, Stuttgart
}%

\begin{abstract}
This letter presents a novel approach for identifying uncorrelated atomic configurations from extensive data sets with a non-standard neural network workflow known as random network distillation (RND) for training machine-learned inter-atomic potentials (MLPs).
This method is coupled with a DFT workflow wherein initial data is generated with cheaper classical methods before only the minimal subset is passed to a more computationally expensive ab initio calculation.
This benefits training not only by reducing the number of expensive DFT calculations required but also by providing a pathway to the use of more accurate quantum mechanical calculations for training.
The method's efficacy is demonstrated by constructing machine-learned inter-atomic potentials for the molten salts KCl and NaCl.
Our RND method allows accurate models to be fit on minimal data sets, as small as 32 configurations, reducing the required structures by at least one order of magnitude compared to alternative methods.
\end{abstract}

\keywords{Machine learning, neural network potentials, random network distillation, graph neural networks.}
\maketitle

Data-driven approaches for reconstructing potential energy surfaces have provided scientists with a unique environment for combining two thriving research areas: machine learning and molecular dynamics.
These machine learning approaches aim to use data from expensive ab initio calculations such as density functional theory (DFT) to fit a model, which may then be used to perform molecular dynamics (MD) simulations at roughly the speed and on scales of a classical approach while retaining the accuracy of the ab initio computations.
The last decade has seen significant advances in the use of machine learning algorithms for the development of these potentials (MLPs), be it Gaussian process regression~\cite{bartok10a}, neural networks~\cite{behler07a, schuett17a, zaverkin20a}, or other kernel methods~\cite{balabin11a, rupp12a}.
A fundamental component to fitting these potentials that has recently become an active area of research is how to select data from these ab initio computations so that one minimises the size of training data sets while maximally representing the underlying potential energy surface (PES).
Typically, this data selection is made uniformly in time, energy, or local energies if a classical potential is used at the initial data selection stages~\cite{cole20a, shao20a, tovey20a, finkbeiner21a}.
In more recent studies, active learning approaches have been implemented to iteratively correct a potential as it ventures into poorly defined areas of configurations space~\cite{sivaraman20a}.
In some cases, configurations are deliberately constructed, such as in the case of RAG sampling~\cite{choi20a} or kernel functions applied to identify unique structures in descriptor space~\cite{de16a}.
With our focus continually on the strictly physical properties of configurations, it can sometimes be instructive to look into methods adopted by the broader machine learning community, which ventures far beyond the realm of molecular dynamics simulations.
One such approach developed in reinforcement learning is Random Network Distillation or RND~\cite{burda18a}.
This approach has been used previously to identify unseen regions of target space for a reinforcement learner and ignore those regions the machine learning algorithm is believed to have explored~\cite{burda18a}.
However, the design of the problem closely mirrors that of selecting data for the development of machine-learned inter-atomic potentials and, therefore, is of interest to the community.
    \begin{figure*}[ht!]
        \centering
        \includegraphics[width=\linewidth]{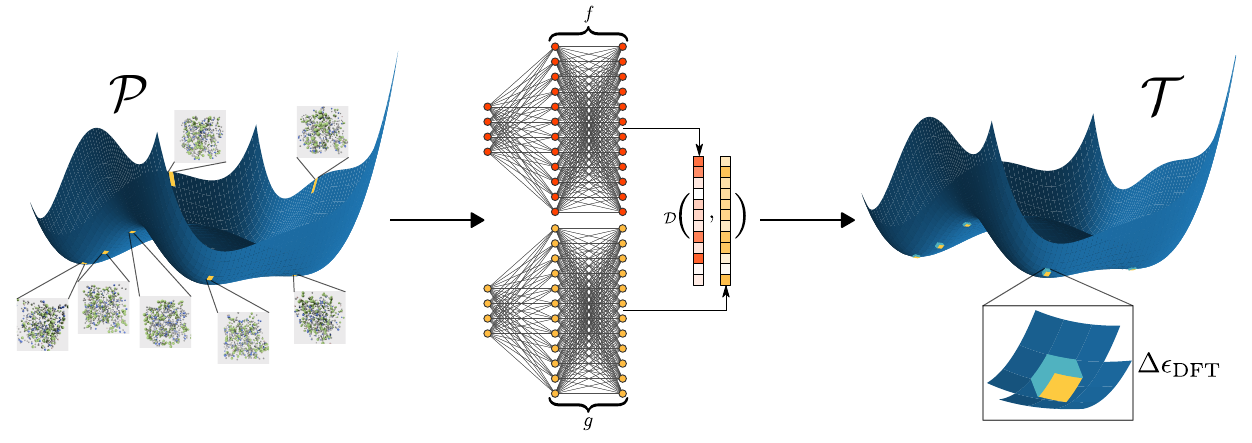}
        \caption{Use of random network distillation to fill a training set $\mathcal{T}$.
        In the initial stage, classical MD simulations are used to sample configuration space and build the data pool $\mathcal{P}$ before the RND architecture is used to select unique configurations and add these to the training data.
        This training data is then passed through a DFT calculation to label the configurations with energy and forces before training a machine-learned potential.
        }
        \label{fig:rnd-outline}
    \end{figure*}
    RND is a method that utilises the intrinsic bias of a neural network architecture to identify regions of the underlying data manifold that will result in a better model after training~\cite{tovey23b}.
    When used for data selection, the goal of RND is to take a large set of data and reduce it to a much smaller but still representative subset on which a model can be trained.
    The method is built upon two neural networks, the target network: $f: \mathcal{R}^{M} \rightarrow \mathcal{R}^{N}$ which acts as an embedding operation for the data, and the predictor network: $g: \mathcal{R}^{M} \rightarrow \mathcal{R}^{N}$ which is trained to predict the output of the target network iteratively.
    Before the data selection occurs, the RND mechanism must be seeded. 
To do so, all points in the large data set are passed through each neural network, and a distance metric is used to compute the distance between the representations generated by $f$ and $g$ for each point.
    The point with the greatest distance, $p_{i}$, is selected and added to the training set, $\mathcal{T}$. 
    The predictor network, $g$, is then trained on the representation generated by $f(p_{i})$.
    This process is continued until a data set of a desired size has been selected.
    
    Our work applies RND to selecting a representative subset of atomistic configurations on which a machine-learned potential will be trained.
    In building the initial data pool from which the subset is selected, large amounts of configuration space must be covered so that the chosen training set is informative.
    One approach is to use classical MD simulations to quickly span the configuration space at a lower accuracy.
In this work, MD simulations are performed in systems made up of 100 atoms in a Nos\`{e}-Hoover chain~\cite{hoover85a, nose84a} enforced NPT ensemble using the LAMMPS simulation software~\cite{plimpton95a}.
    Interactions between the constituent atoms are defined using the Born-Meyer-Huggins-Tosi-Fumi potential~\cite{tosi64a, fumi64a, mayer33a, born32a, huggins33a} parameterized based on literature values~\cite{pan16a} and accompanied by PPPM electrostatic corrections~\cite{eastwood80a}.
    The simulations are run under a temperature ramp from \si{1100}{K} and \si{1700}{K} to cover the liquid phase of the salts.
    From this data pool, RND selects representative subsets of varying sizes.
    For the application of RND, atomic configurations are mapped into a descriptor space using untrained SchNet graph-based representations~\cite{schuett17a, schuett18a}.
    These representations are then passed through the target and predictor network to perform the data-set selection.
    Once a subset is selected, single-point density functional theory (DFT) calculations are performed on the smaller data sets.
    These DFT simulations are performed with the CP2K simulation software~\cite{kuehne20a}, using the PBE-GGA~\cite{perdew96a} functionals, double-zeta MOLOPT basis sets optimized for dense liquids~\cite{vandevondele07a}, GTH pseudo-potentials~\cite{goedecker96a}, and RVV10 non local integral corrections~\cite{sabatini13a}.
    The workflow from classical MD to DFT single-point calculations is outlined in Figure~\ref{fig:rnd-outline}.
    While this classical to ab initio transfer method appears to work in the case of simple liquids, it relies on the similarity of the configuration spaces across these scales. Therefore, it is not a priori valid for more complex systems, and further investigation should be performed in this direction.
    A benefit of RND as a data-selection method is that it scales only with N data points desired in the final data set, as the use of two neural networks introduces some concept of memory of what has been seen before, thus avoiding the expensive nature of other descriptor-based selection methods.
    Furthermore, it separates itself from other descriptor-based methods in that it requires no training in the SchNet representation beforehand. Therefore, it is agnostic to the descriptor and imposes little to no bias on the problem.

\begin{figure*}[ht!]
       \begin{center}
          \renewcommand\sffamily{}
          \hspace*{-0.6cm}
          \input{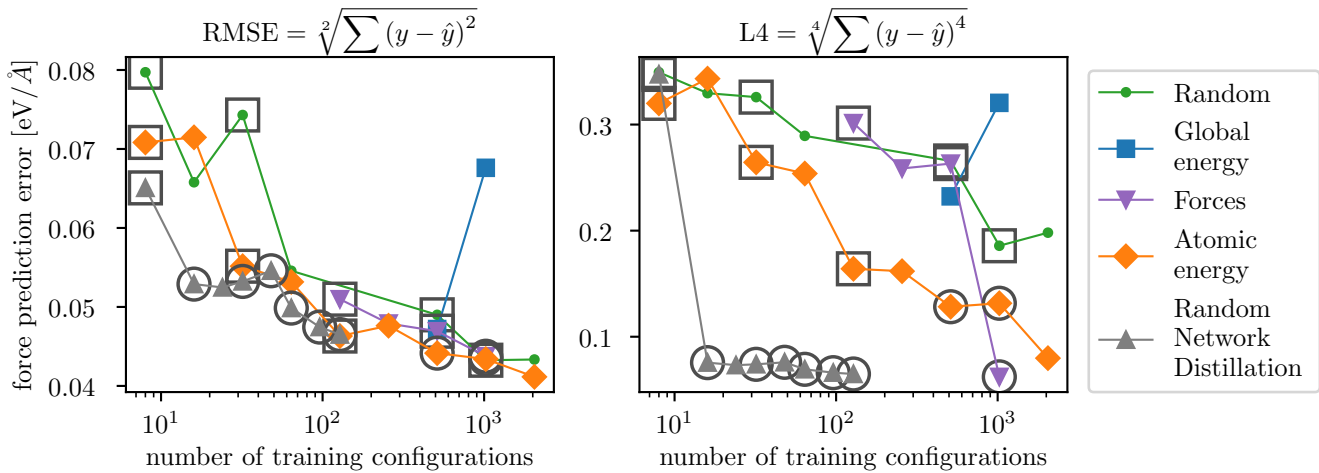}
          \vspace*{-1.cm}
       \end{center}
       \caption{
       The RMSE and L4 loss compared to the number of training configurations for different data selection algorithms shows the convergence of the model loss with respect to the number of training configurations used.
       Circles correspond to those models that could be used to run a stable MD simulation, whereas a square indicates that the potential failed when deployed in a simulation.
       This labelling measures how well the training data represents the configuration space.
                }
       \label{fig:errors}
    \end{figure*}

    After selecting the subsets, machine learning models are trained on the ab initio data.
    This work uses the machine learning framework SchNet~\cite{schuett17a, schuett18a}.
    SchNet is a graph neural network (GNN) based architecture that builds representations from atomic coordinates while respecting the symmetries inherent to the system.
    Models are trained on subsets of varying sizes and compared with more commonly used training data selection methods.
    Figure~\ref{fig:errors} outlines the results of the investigation.
    The figure displays both the RMSE and L4 error calculations for the force predictions of the machine learning models on previously unseen validation data as a function of data-set size for the KCl model (see SI for NaCl plots).
    In each plot, the colour and shape of the lines correspond to a data-selection method, black circles surrounding a point symbolise that a successful MD simulation was performed using this model, and a black square shows that the simulation failed before 100 ps.
    Simulation failure is decided by either drift in energy and temperature, artifacts in the radial distribution function computations, or large forces experienced during the run.
    In the RMSE plots, while it is clear that RND generates models with lower loss values, the differences are not large compared with the other techniques.
    What is clear is that far more of the RND-trained models can perform MD simulations, as seen in the number of circles along the line.
    This trend is elucidated in the L4 error plot, where we can see that the RND-trained data sets converge much faster than all other methods to a minimum value.
    L4 error values have the impact of penalising outliers to a greater extent than their RMSE counterparts.
    The reduction in L4 error suggests that RND can identify maximally separated points, thus reducing the number of outliers in the validation data.
    This trend persists even when compared with other data selection techniques, which explicitly consider local atomic effects, e.g., force selection and atomic energy selection.
    Interestingly, the L4 error coincides with the successful running of a simulation. 
    This relationship suggests that using loss functions that penalise outliers significantly is a good indicator of whether a potential will succeed.

\begin{figure}
        \begin{center}
          \renewcommand\sffamily{}
          \hspace*{-.65cm}
          \input{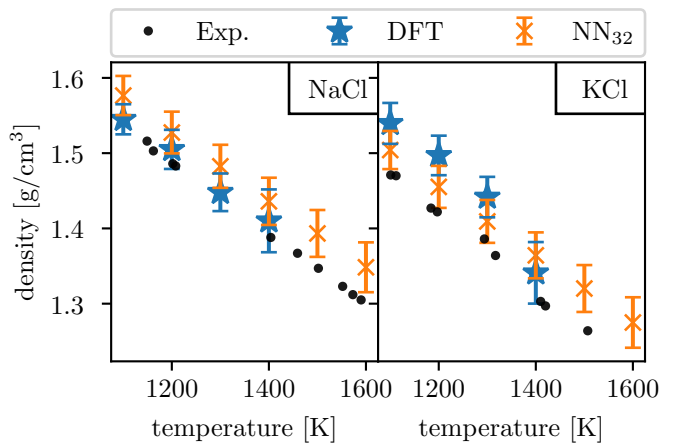}
          \vspace*{-1.0cm}
        \end{center}
        \caption{Density of each salt at different temperatures computed with the machine learned potentials trained on 32 configurations (orange crosses), using pure DFT-MD (blue stars) and experimental data (black dots) taken from Ref \cite{kirschbaum62a}.}
        \label{fig:density_plot}
    \end{figure}

    With successful model fits, the trained potentials can be utilised in scaled-up MD simulations to measure relevant properties.
    One such thermophysical observable of interest to the community is the density of a liquid at different temperatures.
    Density is typically challenging for machine-learned potentials to reproduce as it requires a good representation of configuration space in the training data, typically achieved through active learning and accurate ab initio data~\cite{wang11a}.
    NPT simulations are performed using a custom-written SchNet plugin for LAMMPS~\cite{plimpton95a} on scaled-up system sizes of 400 atoms.
    Densities are computed from 1 ns simulations at several temperatures and plotted against DFT and experimental density values in Figure~\ref{fig:density_plot}.
    The DFT values are taken from 10 ps DFT-MD simulations in an NPT ensemble with 400 atoms using the same DFT parameters as in the single-point calculations.
    We can see that the MLPs accurately reproduce the underlying DFT data with temperature, suggesting that the RND-selected data set of only 32 configurations adequately mapped the configuration space of the salts.

\begin{figure}
        \begin{center}
          \renewcommand\sffamily{}
          \hspace*{-.65cm}
          \input{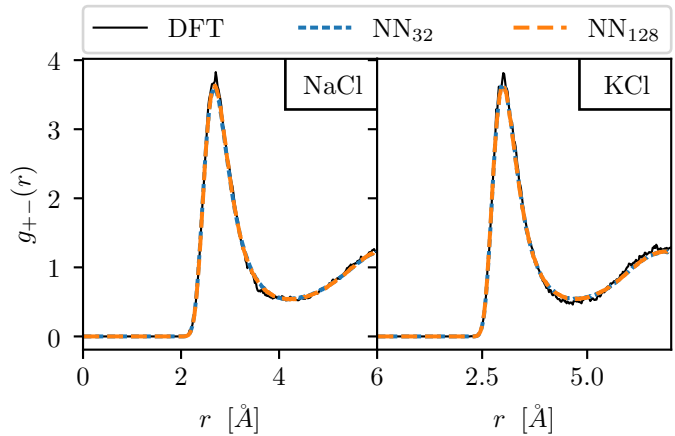}
          \vspace*{-1.0cm}
        \end{center}
        \caption{Comparison of radial distribution functions generated from MD simulations performed in an NVT ensemble using the machine learned potentials and the underlying density functional theory data.}
        \label{fig:rdfs}
    \end{figure}

    Another important observable in MD simulations is the radial distribution function (RDF), which can be directly related to the DFT data on which the ML model was trained.
    To generate data for the RDF calculations, NVT simulations are performed at densities fixed to those of the compared experimental values.
    The MDSuite post-processing software~\cite{tovey23a} is then used to compute the RDFs.
    The MD simulations are run for 1 ns using a Nos\`{e}-Hoover chain~\cite{hoover85a, nose84a} with a coupling constant of 100 fs.
    To create reference data, 10 ps DFT-MD runs in an NVT ensemble are also performed using the parameters described for the single-point calculations.
    Figure~\ref{fig:rdfs} compares the anion-cation RDF curves for the machine-learned potentials against the reference DFT data. 
RDFs are shown for two different models trained on different amounts of data.
    In all cases, the ML potentials accurately reproduced the underlying DFT data.

    Finally, the dynamic properties of the salts are assessed in the form of self-diffusion coefficients and ionic conductivity.
    The trajectories from 1 ns MD studies are used along with the MDSuite software~\cite{tovey23a} in the computation of the properties.
    Tables~\ref{tab:diffusion-values} and~\ref{tab:ionic-conductivity} compare the results computed from the MD simulations with those of the experiment.
        \begin{table}[]
        \begin{tabular}{lccl}
        \hline
                          & Species & D$_{\text{sim}}$  & \multicolumn{1}{c}{D$_{\text{exp}}$} \\ \hline
        \multirow{2}{*}{NaCl} & Na      & $1.118 \pm 0.006$ & $1.052 \pm 0.210$                    \\
                          & Cl      & $0.903 \pm 0.005$ & $0.842 \pm 0.168$                    \\ \hline
        \multirow{2}{*}{KCl}  & K       & $1.052 \pm 0.005$ & $1.005 \pm 0.201$                    \\
                          & Cl      & $1.069 \pm 0.006$ & $0.905 \pm 0.181$                    \\ \hline
        \end{tabular}
        \caption{Self-diffusion coefficients computed from the ML potential simulations compared with experimental fits from Ref~\cite{janz82a}}
        \label{tab:diffusion-values}
        \end{table}
        
        \begin{table}[ht]
        \begin{tabular}{lll}
        \hline
        & \multicolumn{1}{c}{$\sigma_{\text{Sim}}$} & \multicolumn{1}{c}{$\sigma_{\text{Exp}}$} \\ \hline
        NaCl        & $3.885 \pm 0.118$                         & $3.954 \pm 0.032$                         \\
        KCl         & $2.779 \pm  0.057 $                       & $2.517 \pm 0.044$                         \\ \hline
        \end{tabular}
        \caption{Ionic conductivity data from the ML potential simulations compared with experimental values taken from Ref~\cite{janz67a}}
        \label{tab:ionic-conductivity}
    \end{table}

    We see that for both salts, the self-diffusion coefficients match well with experimental values, suggesting an accurate MLP trained on good ab initio data.
    Ionic conductivity measurements are also in good agreement with experimental values.
    
    We have demonstrated that random network distillation can be used to identify relevant atomic configurations to train data-driven inter-atomic potentials.
    We did so by fitting machine-learned potentials on systems of NaCl and KCl using the SchNet framework.
    Furthermore, our data selection method outperformed several other approaches, including global energy selection, local energy selection, and force-based selection in model convergence.
    We have performed molecular dynamics simulations on scaled systems of up to 500 ion pairs and for more than 1 ns to validate the ML potentials on more significant length and time scales.
    The structural and dynamic properties computed from these simulations were shown to reproduce pure ab initio investigations and experimental data adequately.
    Finally, we showed that RND is capable, without additional active learning, of performing stable NPT simulations and converging to the system density expected from DFT.
    These results support several conclusions.
    Random network distillation is an efficient method for identifying unique configurations for training MLPs.
    Single-point DFT calculations on classically generated configurations are sufficient for producing accurate training data for machine learning models.
    At least for chemically simple systems, the number of configurations required for an NPT-capable model yielding accurate structures, dynamics, and densities is significantly smaller than previously reported in the literature, resulting in improved training time and reduced computational demand.
    This minimal training set also provides an avenue for extending the potentials to higher level ab-initio calculations such as coupled cluster~\cite{coester60a} or configuration interaction~\cite{sherrill99a} and thereby producing MLPs beyond the accuracy of DFT.
    Future work should investigate the application of RND to more complex systems and better understand its limitations.
    
\begin{acknowledgments}
The authors acknowledge financial support from the German Funding Agency (Deutsche Forschungsgemeinschaft DFG) under Germany's Excellence Strategy EXC 2075-390740016.
This work was supported by SPP 2363- "Utilization and Development of Machine Learning for Molecular Applications – Molecular Machine Learning."
Funded by the Deutsche Forschungsgemeinschaft (DFG, German Research Foundation), Project-No 497249646
\end{acknowledgments}~\nocite{*}

\newpage
\bibliography{main}

\newpage
\appendix

\section{Detailed Results}\label{sec:detailed_results}
\subsection{Model Performance}\label{subsec:model_performance_detailed}
\begin{figure*}
    \begin{center}
       \renewcommand\sffamily{}
       \input{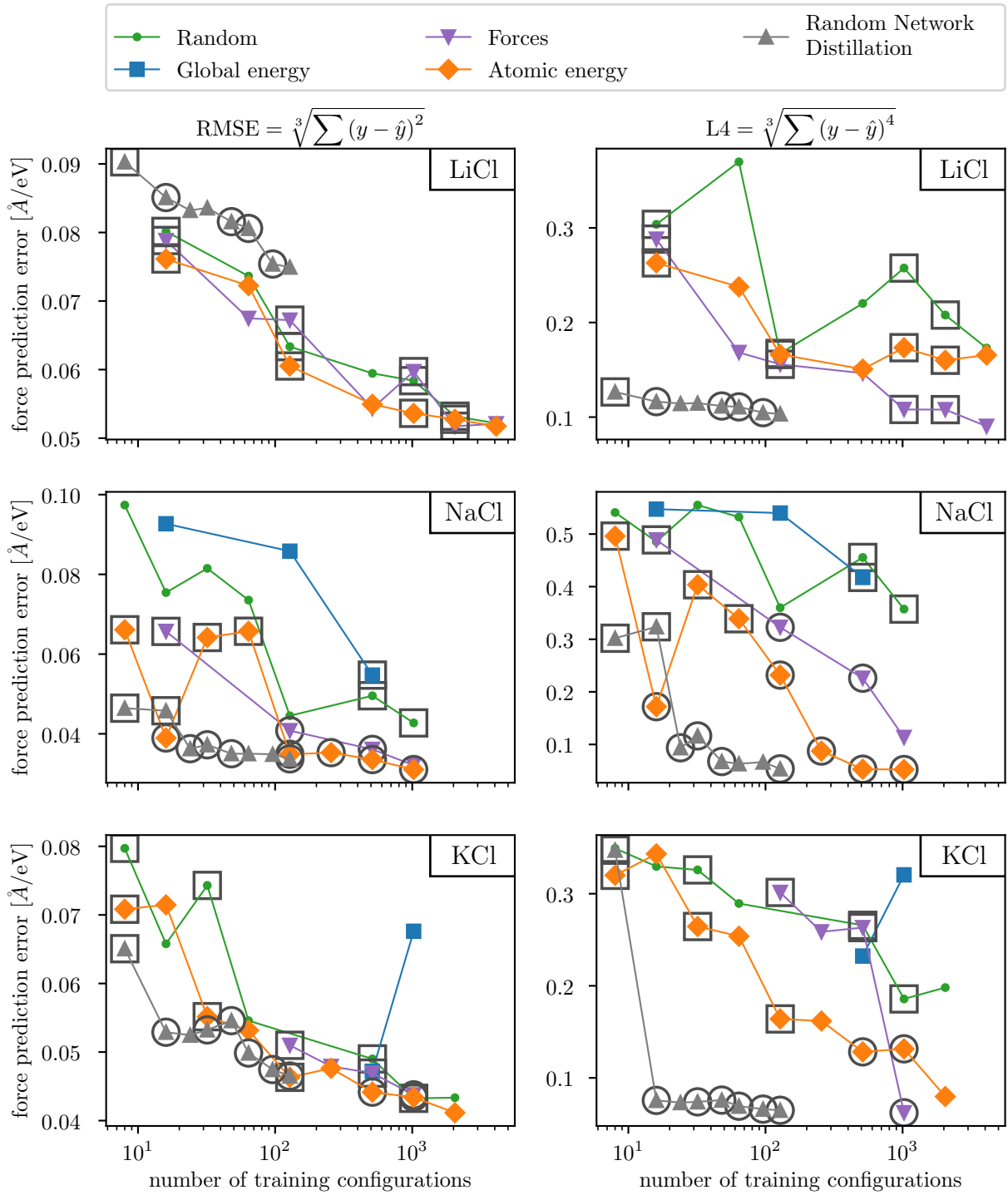}
       \vspace*{-1.cm}
    \end{center}
    \caption{RMSE and L4 loss compared to the number of training configurations for different data selection algorithms showing the convergence of the model loss with respect to the number of training configurations used.
 Circles correspond to those models that could be used to run a stable MD simulation.}
    \label{fig:si_errors}
 \end{figure*}

Model performance has been evaluated using the L4-norm as well as the RMSE.
Figure~\ref{fig:si_errors} displays the results of this evaluation for all of the
salts studied.

 \subsection{Structural Properties}\label{subsec:structural_properties_detailed}
Here, the radial distribution functions of the neural network potential generated trajectories
are compared with ab initio data for all salts studied.
 \begin{figure*}
    \begin{center}
       \renewcommand\sffamily{}
       \input{figures/si_figure_2.pgf}
       \vspace*{-0.5cm}
    \end{center}
    \caption{comparison of radial distribution functions generated from MD simulations performed in an NVT ensemble using the machine learned potentials and the underlying density functional theory data.}
    \label{fig:si_rdfs}
 \end{figure*}

\subsection{Dynamic Properties}\label{subsec:dynamic_properties_detailed}
Dynamic quantities are defined here as those computed from time dependence within the
MD trajectory.
In our case, this includes the diffusion coefficients and ionic conductivity, as they are both
computed from correlations in time.
Auto-correlation functions have been computed for several ranges and integrated over different regions along the computed curve to calculate the coefficients.
This approach allows for a confident measure of the transport property.
\begin{figure*}
    \begin{center}
       \renewcommand\sffamily{}
       \input{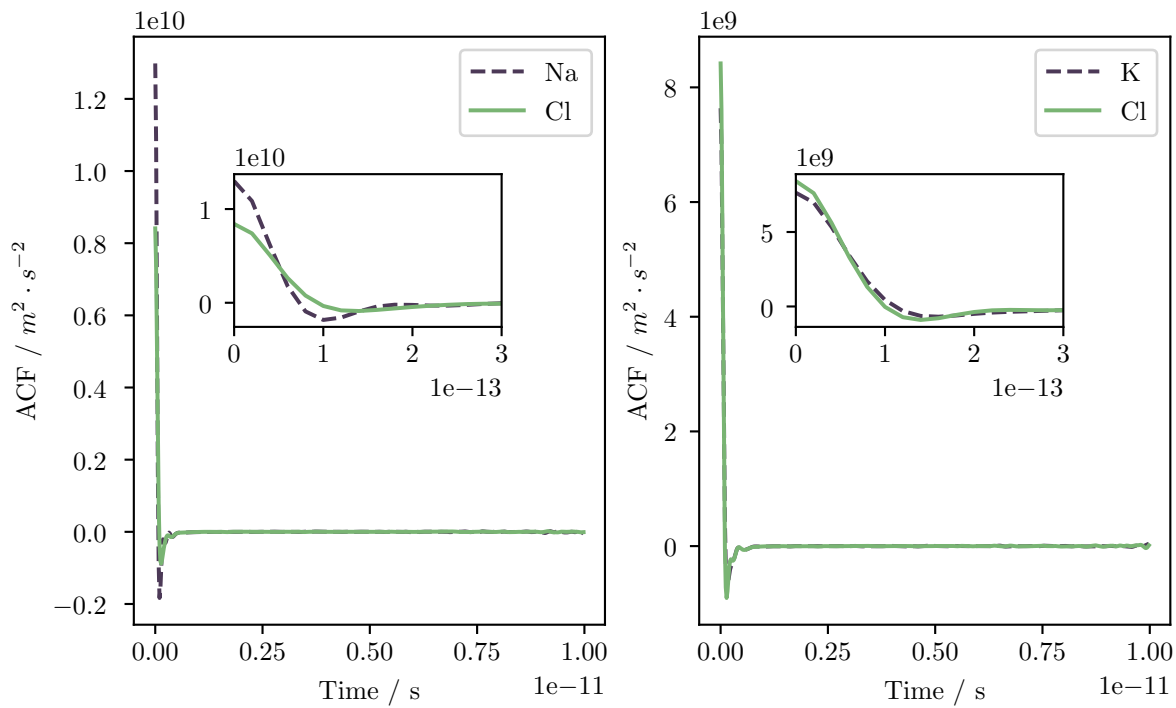}
       \vspace*{-0.5cm}
    \end{center}
    \caption{Velocity autocorrelation functions for the computation of the diffusion coefficients.
During the fitting process, several of these functions are produced over different data ranges and
integration ranges.
    }
    \label{fig:vacfs}
\end{figure*}

\begin{figure*}
    \begin{center}
       \renewcommand\sffamily{}
       \input{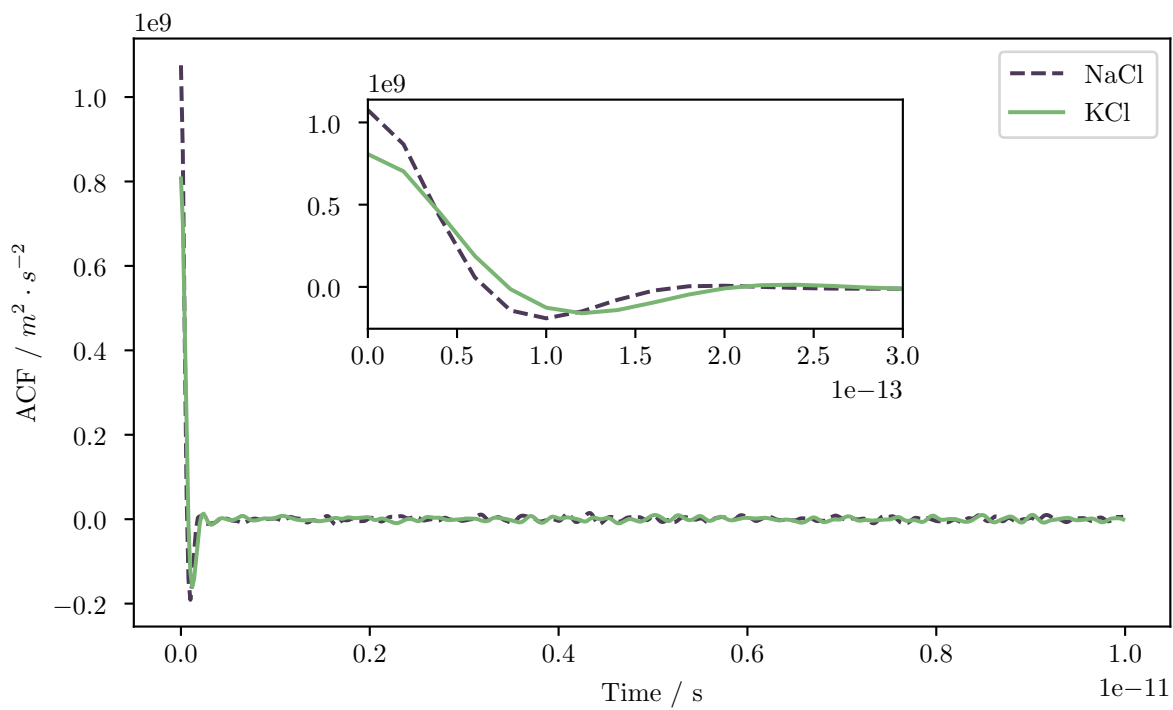}
       \vspace*{-0.5cm}
    \end{center}
    \caption{Current autocorrelation functions for the computation of the ionic conductivity.
During the fitting process, several of these functions are produced over different data ranges and
integration ranges.
    }
    \label{fig:jacfs}
\end{figure*}

\end{document}